 \documentclass[pre,twocolumn,twoside,a4paper,floatfix]{revtex4}
 \usepackage{graphicx}
 \usepackage{amssymb} 
\begin{document}

 \title{Reaching large lengths and long times in polymer dynamics
simulations}

 \author{A. van Heukelum} 
 \affiliation{Institute for Theoretical Physics, Utrecht University,\\
Leuvenlaan 4, 3584 CE Utrecht, The Netherlands}

 \author{G. T. Barkema} 
 \affiliation{Institute for Theoretical Physics, Utrecht University,\\
Leuvenlaan 4, 3584 CE Utrecht, The Netherlands}


 \begin{abstract}
   A lattice model is presented for the simulation of dynamics in
polymeric systems. Each polymer is represented as a chain of monomers,
residing on a sequence of nearest-neighbor sites of a
face-centered-cubic lattice. The polymers are self- and mutually
avoiding walks: no lattice site is visited by more than one polymer, nor
revisited by the same polymer after leaving it. The dynamics occurs
through single-monomer displacements over one lattice spacing. To
demonstrate the high computational efficiency of the model, we simulate
a dense binary polymer mixture with repelling nearest-neighbor
interactions between the two types of polymers, and observe the phase
separation over a long period of time. The simulations consist of a
total of $46,080$ polymers, $100$ monomers each, on a lattice with
$13,824,000$ sites, and an interaction strength of $0.1 k_B T$. In the
final two decades of time, the domain-growth is found to be $d(t) \sim
t^{1/3}$, as expected for a so-called ``Model~B'' system.
 \end{abstract}

 \maketitle

 \section{Introduction}

   Lattice models are widely used as a theoretic tool to study polymer
solutions. Most notably, Flory~\cite{flory1942, flory1944} and
Huggins~\cite{huggins1942a, huggins1942b} introduced a mean-field theory
of a particular simple model which has proven powerful enough to explain
solubility properties and liquid-liquid phase separation of polymer
mixtures. An extension to the model, describing the dynamics of such
systems, has been introduced by Cahn and
Hilliard~\cite{cahn1958,cahn1959} and Hillert~\cite{hillert1961}. Some
of the earliest computer simulations of lattice polymers were performed
in 1962 by Verdier and Stockmayer~\cite{verdier1962}, who studied the
dynamics of polymers, modeled by self-avoiding walks on square and cubic
lattices. Kolinski {\em et al.}~\cite{kolinski1987a,kolinski1987b} added
two-monomer moves to the dynamics and simulated a homopolymeric melt.

   To describe the polymer dynamics in a melt with more detail than the
random-walk approach, Carmesin introduced in 1988 the bond fluctuation
model~\cite{carmesin1988}. In the two-dimensional version of this model,
each monomer occupies four ($2 \times 2$) lattice sites of a square
lattice. Multiple occupation of lattice sites is not permitted. Monomers
adjacent in the chain are connected by bonds with lengths between 2 and
$\sqrt{13}$. The dynamics of the polymer chain consists of the
displacement of single monomers to nearest-neighbor lattice sites,
restricted by the constraints on bond length and excluded volume. The
bond-fluctuation model was extended to three dimensions in 1991, by
Deutsch and Binder~\cite{deutsch1991}. This model has been used by
several groups~\cite{paul1991, kreer2001} to simulate dense polymer
melts.

   In 1971, de Gennes~\cite{degennes1971} proposed that the main
mechanism of polymer dynamics in gels and dense polymer mixtures is {\em
reptation}, i.e., diffusion of stored length. To verify the theoretical
predictions of de Gennes, Evans and Edwards~\cite{evans1981} introduced
the {\it cage model}. in this model, a single polymer is simulated as a
random walk on a square or cubic lattice. In the limit of tight gels,
the dynamics reduces to the re-orientation of pairs of connected
segments pointing in opposite direction (``kinks''). The cage model has
been used for the study of diffusion and relaxation times of a single
polymer in a gel~\cite{evans1981, deutsch1989a, deutsch1989b,
barkema1998}, as well as for star polymers~\cite{barkema1999}. Extending
the model to include excluded-volume effects or interactions between
polymers is not easy.

\begin{figure}
 \includegraphics[width=3.25in]{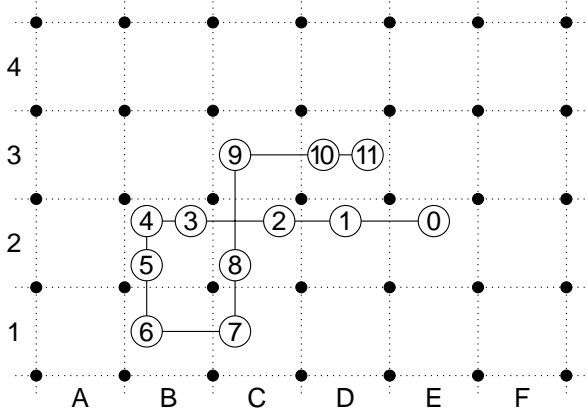}
 \caption{The repton model on a square lattice. The monomers in the
polymer chain are connected by bonds of zero or unit length and each
monomer is assigned to a face of the lattice, such that adjacent
monomers in the chain occupy either the same or nearest-neighbor lattice
sites.  If a monomer is connected to one neighbor by a bond with zero
length, and to its other neighbor by a bond with unit length, it can
join its other neighbor. By doing so, the bond with zero length---a
representation of ``stored length''---diffuses along the chain. For
instance, in the configuration depicted here, monomer 5 can join its
neighbor 6. The ends of the polymer allow fluctuations of the total
amount of stored length.}
 \label{fig:repton}
 \end{figure}

   An alternative model for a single reptating polymer was proposed by
Rubinstein~\cite{rubinstein1987}. In this model, known as the {\em
repton model}, the polymer is represented as a chain of monomers,
residing on the sites of a square or cubic lattice. Monomers adjacent
along the chain reside in either nearest-neighbor lattice sites, or in
the same site. The latter case corresponds to the presence of stored
length, a central concept in de Gennes' description of entangled polymer
dynamics. The dynamics of the polymer is strictly limited to reptation:
the only kind of moves allowed to a monomer in the interior of the chain
is that it can move to an adjacent lattice site, provided that one of
its adjacent neighbors is already on the site to which it is moving, and
the other is in the site it leaves. Per unit of time, each monomer
attempts to move in each direction statistically once. The repton model
is illustrated in Figure~\ref{fig:repton}.
 The repton model has been studied extensively by one of
us~\cite{newman1997, barkema1997}.

   The repton model can be extended to simulate polymers with
self-avoiding walk statistics, as well as many interacting polymers;
this is the route followed in this manuscript. In the next section, we
discuss the extensions to the repton model which are required to
describe equilibrium and dynamical properties of dense polymer solutions
and melts. Since the dynamics of polymeric systems occurs on long
time-scales and large length-scales, computational efficiency is of
paramount importance. We therefore discuss in detail how the resulting
extended repton model can be implemented with high computational
efficiency, exploiting a technique known as multispin coding. At the end
of this manuscript, we show the usefulness of our work by presenting
simulation results on the phase separation of a dense binary polymer
mixture.

 \section{Extensions of the repton model to simulate polymer melts}

   The repton model as proposed by Rubinstein simulates a single polymer
obeying random-walk statistics, with dynamics limited to the diffusion
of stored length. Two extensions to the repton model are needed to use
it for the simulation of entangled chains: the chains have to obey
volume exclusion and on the other hand, the constraint that disallows
sideways motion has to be lifted.

   The repton model describes ``stored length'' as two or more monomers
on the same lattice site. Therefore, we cannot limit the polymers to
self-avoiding-walk statistics by disallowing two monomers to occupy the
same lattice site, without destroying the reptation dynamics that so
successfully described polymer-diffusion in a gel. As reptation is also
expected to play an important role in the melt, we should keep this kind
of dynamics. The solution to this problem is to limit only the contours
of the polymers to self-avoiding-walk statistics. Thus, multiple
occupation of a lattice site is allowed only for two or more monomers
which are adjacent along the chain. One convenient side-effect of this
choice is that since the reptation moves in the interior of the chain do
not affect the tube, they never cause a violation of volume exclusion.

   Sideways motion is implemented by also allowing monomers to move by a
single lattice spacing if this changes the tube of the polymer.
Single-monomer moves work well on lattice structures that contain loops
of three sites, like the triangular and face-centered-cubic (FCC)
lattice; on lattices without such three-site loops, like the square and
cubic lattices, the sideways movement can lead to lots of unwanted
artifacts or, alternatively, must be implemented by allowing bonds to
extend to next-nearest neighbors.

   The repton model is often studied in its projected version, in which
the state of the polymer is characterized by the bonds $s_i$, $i=1\dots
N-1$, between the monomers $i=1\dots N$, where $s_i\equiv x_{i+1}-x_i$
is equal to zero or $\pm 1$. Usually all allowed moves are attempted
with the same (unit) rate, with the consequence that the three values
for $s_i$ occur with equal probability. The density of stored length
(probability that $s_i$ is zero) is then 1/3. To obtain the same density
of stored length in higher-dimensional repton models (before
projection), the rates for moves which decrease the stored length should
equal $2/z$ times those of the reverse moves, in which $z$ is the
lattice coordination number: 4, 6 and 12 for the square, cubic or
triangular, and FCC lattices, respectively. To conserve equilibrium
properties, sideways moves which decrease the density of stored length
should also be attempted with rates which are $2/z$ times the reverse
moves.

\begin{figure}
 \includegraphics[width=3.25in]{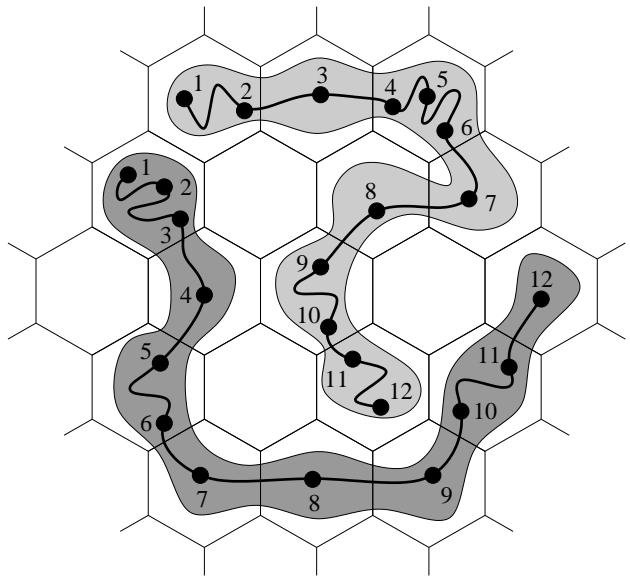}
 \caption{Extended repton model in two dimensions. Sideways motion of a
single monomer, preserving the constraint that bonds are either
zero-length or unit-length, is only possible on lattices that contain
loops of three sites, i.e., triangular (FCC) instead of the more usual
square (cubic) lattice. The excluded volume constraint is implemented by
allowing only adjacent monomers in a polymer to occupy the same lattice
site. Phase separation is induced by nearest-neighbor lattice site
interactions.}
 \label{fig:manymodel}
 \end{figure}

   Figure~\ref{fig:manymodel}
 shows model-polymers on a triangular lattice. In the upper polymer in
the figure, reptation moves are possible for monomers 2, 4, 6, 9, 10,
and 11; sideways moves can be made by monomers 2, 4, 6, 8, 9, 10, and
11. Monomers 3, 5, and 7 cannot move. The end-monomers 1 and 12 can move
to any empty nearest-neighbor site. In the lower polymer, reptation
moves are possible for monomers 3, 5, 6, 10, and 11; sideways moves can
be made by monomers 3, 4, 5, 6, 7, 10, and 11. Monomers 2, 8, and 9
cannot move. The end-monomer 1 can move to any empty nearest-neighbor
site, while end-monomer 12 can only move to the site occupied by monomer
11.

 \section{multispin coding}

   Multispin coding is a programming technique which makes use of the
low-level bit manipulation functions provided by the CPU of our
computer, effectively doing many simple computations in parallel. These
functions include logical operations like AND ($\land$), OR ($\lor$)
exclusive-or (XOR, $\oplus$), negation (NOT, $\lnot$), as well as
bitwise shifts ($\ll$, integer multiplication by a power of 2; $\gg$,
integer division by a power of 2); and arithmetic operations like
addition and multiplication. For a general introduction to multispin
coding we refer to Ref.~\cite{NewmanBarkema99}, Chapter 15. Here, we
will explain this technique analogous to Ref.~\cite{barkema1994}, as
applied to the projected repton model. Next we will discuss the
implementation of this technique in our model for polymer melts.

 \subsection{repton model}

   In the projected repton model, the state of the chain with monomers
$i = 1 \dots N$ is specified by the set of their coordinates $\{x_1,
\dots, x_N\}$. Monomers which are adjacent along the chain have to
reside in either the same lattice site, or nearest neighbors, with the
consequence that the step $s_i \equiv x_{i+1} - x_i$ can only take the
values $\pm 1$ or 0. An alternative way to store the polymer state is
therefore to store the coordinate $x_1$ of the first monomer, and to
store for step $i = 1 \dots N-1$ two arrays of bits $(A_i, B_i)$ that
can take the combination $(0, 0)$ if $s_i = 0$, $(0, 1)$ if $s_i = 1$,
and $(1, 0)$ if $s_i = -1$.

   In one unit of time, each monomer attempts to move in each direction
statistically once. Thus, $2N$ elementary moves, i.e. one monomer
attempting to move in one direction, should be attempted per unit of
time.

   It is a property of the repton model, that if monomer $i$ in the
interior of the chain can join its neighbor $i+1$, it cannot join its
neighbor $i-1$, and vice versa. We exploit this by attempting these two
elementary moves simultaneously. The first and last monomer can move to
both directions, if they are located in the same site as their neighbor
along the chain, so this trick does not apply for these two monomers;
they are therefore selected with a double probability.

   In detail, each monomer in the interior of the chain is selected with a
probability $1/(N + 2)$. If the selected monomer $i$ is located in the
same site as exactly one of its two adjacent neighbors, it will join the
other neighbor. The bits that contain information on whether a move is
possible or not, are $(A_i, B_i)$ and $(A_j, B_j)$ with $j = i - 1$. A
move is possible if one of the two pairs is $(0, 0)$ and the other is
not. If the move is carried out, $(A_i, B_i)$ and $(A_j, B_j)$ are
exchanged. A sequence of logical operations that achieves this is:
 \begin{eqnarray}
 \begin{array}{rcl}
 y & = & (A_i \lor B_i) \oplus (A_j \lor B_j) \\
 m_A & = & (A_i \oplus A_j) \land y \\
 m_B & = & (B_i \oplus B_j) \land y \\
 A'_i & = & A_i \oplus m_A \\
 B'_i & = & B_i \oplus m_B \\
 A'_j & = & A_j \oplus m_A \\
 B'_j & = & B_j \oplus m_B
 \end{array}
 \end{eqnarray}

   The first and last monomers are selected, each, with a probability
$2/(N + 2)$; twice the probability to select a specific monomer in the
interior of the chain. After the selection of an end monomer, the
intended direction is also randomly selected, with 50\% probability to
be in the positive or negative direction. The rates at which elementary
moves are attempted is thus equal for interior monomers and those at the
ends. If the first or last monomer is selected, similar statements
suffice to update $(A_1, B_1)$ or $(A_{N-1}, B_{N-1})$, respectively.
The following statements can be used to update the end-monomer 1, trying
to displace it in the negative or positive direction based on the value
of $r_1$.
 \begin{eqnarray}
 \begin{array}{rcl}
 A'_1 & = & \lnot (B_1 \lor r_1) \\
 B'_1 & = & (\lnot A_1) \land r_1
 \end{array}
 \end{eqnarray}
 However, in case the first monomer is selected, also its coordinate
$x_1$ needs to be updated. Since this coordinate is an integer number
that can take a wide range of values, its update is not implemented in a
multispin coding fashion. Luckily, the probability to select the first
monomer decreases with increasing polymer length.

   The motivation for multispin coding lies in its efficiency. The
simulation of the dynamics of 64 polymers involves $128N$ elementary
moves per time unit. With multispin coding, it involves that the above
sequence of logical operations for interior monomers should be performed
statistically $N-2$ times, which involves only 11 logical operations, 4
loads, and 4 stores. On a fast workstation, this takes 41~ns
of cpu-time, or 0.32~ns per elementary move. For updating the last
monomers, even less logical operations suffice. For the first monomers,
however, the update of the 64 values of $x_1$ cannot be achieved as
efficiently, and a loop over all 64 polymers is inevitable, with the
consequence that the simulation of the first monomer requires 5.8~ns
cpu-time per elementary move.

   The multispin implementation outlined above will thus perform 64
simulations in parallel. These simulations are however correlated, since
they share the sequence of selected monomers. In fact, if at some point
two simulations are in identical polymer configurations, they will stay
in identical configurations ever after; in long simulations of small
systems, this ``locking'' will inevitably happen. Complete locking is
avoided by an uncorrelated choice in the directions in which end
monomers attempt to move, using a 64-bit random bit pattern rather than
a binary choice between all-up or all-down. If desired, more de-locking
can be obtained at the expense of a lower efficiency, by rejecting a
fraction of the allowed moves, also using a 64-bit random bit pattern.
An important remark is, however, that as long as each of the 64
simulations are correct in themselves, one obtains 64 unbiased results;
one should just be careful to assign significance to the spread in those
64 simulations.

 \subsection{model for polymer melts}

   While monomers in the projected repton model live on a
one-dimensional lattice, the monomers in the model that we propose for
polymer melts live on a FCC lattice. It is helpful to note that the
three-dimensional hyperplane, located in a four-dimensional hypercubic
space through the origin and with perpendicular vector $(1, 1, 1, 1)$,
is such an FCC lattice. Stated differently, the set of points $\vec{x} =
(a, b, c, d)$ with integer-valued coordinates, constrained to $a + b + c
+ d = 0$, forms a FCC lattice. The twelve vectors pointing to
nearest-neighbor sites are $\hat{t} = (-1, 1, 0, 0)$, $\hat{u} = (0, -1,
1, 0)$, $\hat{v} = (0, 0, -1, 1)$, $\hat{w} = -(\hat{t} + \hat{u} +
\hat{v})=(1, 0, 0, -1)$, and some of their combinations, as listed in
Table~\ref{tab:nn-vectors}.
 \begin{table}
 \caption{The twelve vectors pointing to nearest neighbors of a FCC
lattice, expressed in combinations of $\hat{t}$, $\hat{u}$, $\hat{v}$
and $\hat{w} = - (\hat{t} + \hat{u} + \hat{v})$.}
 \label{tab:nn-vectors}
 \begin{tabular}{|r@{}r@{}rr@{}rr@{}rr@{}l
      cc@{\,}c@{\,}c@{\,}c@{\,}c|
      r@{}r@{}rr@{}rr@{}l|}
 \hline
 \multicolumn{15}{|c|}{vector in $\mathbb{Z}^4$} &
      \multicolumn{7}{|c|}{$\mathbb{Z}^3$} \\
 \hline
 $($&$-1$&,&$ 1$&,&$ 0$&,&$ 0$&$)$ &
      $\equiv$ &
      $\hat{t}$ & & & & &
      $($&$-1$&,&$ 1$&,&$ 0$&$)$ \\
 $($&$ 0$&,&$-1$&,&$ 1$&,&$ 0$&$)$ &
      $\equiv$ &
      $\hat{u}$ & & & & &
      $($&$ 0$&,&$-1$&,&$ 1$&$)$ \\
 $($&$ 0$&,&$ 0$&,&$-1$&,&$ 1$&$)$ &
      $\equiv$ &
      $\hat{v}$ & & & & &
      $($&$ 1$&,&$ 1$&,&$ 0$&$)$ \\
 $($&$ 1$&,&$ 0$&,&$ 0$&,&$-1$&$)$ &
      $\equiv$ &
      $\hat{w}$ & & & & &
      $($&$ 0$&,&$-1$&,&$-1$&$)$ \\
 $($&$-1$&,&$ 0$&,&$ 1$&,&$ 0$&$)$ &
      = &
      $\hat{t}$&$+$&$\hat{u}$ & & &
      $($&$-1$&,&$ 0$&,&$ 1$&$)$ \\
 $($&$ 0$&,&$ 1$&,&$ 0$&,&$-1$&$)$ &
      = &
      $\hat{t}$&$+$&$\hat{w}$ & & &
      $($&$-1$&,&$ 0$&,&$-1$&$)$ \\
 $($&$ 0$&,&$-1$&,&$ 0$&,&$ 1$&$)$ &
      = &
      $\hat{u}$&$+$&$\hat{v}$ & & &
      $($&$ 1$&,&$ 0$&,&$ 1$&$)$ \\
 $($&$ 1$&,&$ 0$&,&$-1$&,&$ 0$&$)$ &
      = &
      $\hat{v}$&$+$&$\hat{w}$ & & &
      $($&$ 1$&,&$ 0$&,&$-1$&$)$ \\
 $($&$-1$&,&$ 0$&,&$ 0$&,&$ 1$&$)$ &
      = &
      $\hat{t}$&$+$&$\hat{u}$&$+$&$\hat{v}$ &
      $($&$ 0$&,&$ 1$&,&$ 1$&$)$ \\
 $($&$ 0$&,&$ 0$&,&$ 1$&,&$-1$&$)$ &
      = &
      $\hat{t}$&$+$&$\hat{u}$&$+$&$\hat{w}$ &
      $($&$-1$&,&$-1$&,&$ 0$&$)$ \\
 $($&$ 0$&,&$ 1$&,&$-1$&,&$ 0$&$)$ &
      = &
      $\hat{t}$&$+$&$\hat{v}$&$+$&$\hat{w}$ &
      $($&$ 0$&,&$ 1$&,&$-1$&$)$ \\
 $($&$ 1$&,&$-1$&,&$ 0$&,&$ 0$&$)$ &
      = &
      $\hat{u}$&$+$&$\hat{v}$&$+$&$\hat{w}$ &
      $($&$ 1$&,&$-1$&,&$ 0$&$)$ \\
 \hline
 \end{tabular}
 \end{table}
 The vector $(0, 0, 0, 0)$ is used as the representation for a
zero-length bond.

   As in the projected repton model, the state of a polymer on our FCC
lattice can be specified by the location $\vec{x}_1$ of the first
monomer, plus additionally the direction in which adjacent neighbors are
located. We choose for the polymer melt model to store the four
directional bits in a single word: bits $k$, $k+16, k+32$ and $k+48$ of
the 64-bit word $D_i$ indicate the vector pointing from monomer $i$ to
monomer $i+1$. Thus, sixteen polymers are updated simultaneously.

   The coordinates of the $i^{\mathrm{th}}$ monomer in polymer $k$ can
then be retrieved by summing over all words $D_i$ bits $k$, $k+16$,
$k+32$ and $k+48$, yielding respectively the numbers $a$, $b$, $c$ and
$d$; the monomer position is then $\vec{x}_i = a\hat{t} + b\hat{u} +
c\hat{v} + d\hat{w}$. Note that these summations require only $3 i$
operations (a right shift of $k$ bits, masking the direction bits, and
adding to the sum), since the summation in the different bits can be
done in a single operation. Since in the polymer melt model we need the
coordinates frequently (every time we attempt a possible sideways move),
and since our polymers are often several hundred monomers long, we do
not keep track explicitly of only the position $\vec{x}_1$ of the first
monomer, but also of the last monomer as well a few other monomers along
the chain, such that the distance along the chain to a monomer with
known position is always less than 15. Of course, to retrieve the
position $\vec{x}_i$, we start from the nearest monomer with known
position in either direction along the polymer.

   The implementation proceeds analogous to the projected repton model.
Also here, if an interior monomer can move in one direction, its move in
the other direction is blocked; this can be exploited as in the repton
model, by combining two elementary moves. The precise operations are:
 \begin{eqnarray}
 \begin{array}{rcl}
 y_0 & = & D_i \lor (D_i \gg 32) \\
 y_1 & = & y_0 \lor (y_0 \gg 16) \\
 z_0 & = & D_j \lor (D_j \gg 32) \\
 z_1 & = & z_0 \lor (z_0 \gg 16) \\
 m_0 & = & (y_1 \oplus z_1) \land (2^{16} - 1) \\
 m_1 & = & (D_i \oplus D_j) \land (M * m_0) \\
 D'_i & = & D_i \oplus m_1 \\
 D'_j & = & D_j \oplus m_1
 \end{array}
 \end{eqnarray}
 Here, $A \ll k$ stands for shifting the word $A$ over $k$ bits to the
left, and $M = 2^{0} + 2^{16} + 2^{32} + 2^{48}$ is a constant, used to
duplicate the low 16 bits in the higher bits of the word. Thus, with
only 15 operations, 2 loads, and 2 stores, we have performed 32
elementary moves. On a fast workstation, the above implementation
requires 1.25~ns cpu-time per elementary move.

   As in the projected repton model, if the first monomers are
displaced, whose positions are tracked, these positions have to be
updated in a loop over the 16 polymers. We succeeded in implementing
moves of the first monomers in 82~ns cpu-time per elementary move. Other
monomers whose position is tracked require roughly the same
computational effort. Displacement of the first and last monomers is
attempted at twice the rate of the other monomers, for the same reason
as in the projected repton model.

   Besides the reptation moves, the dynamics consists of sideways moves.
If a sideways move is attempted on monomer $i$, the relevant bit
patterns are those indicating the direction from monomer $i-1$ to $i$
and from $i$ to $i+1$. These bit patterns, as listed in
Table~\ref{tab:nn-vectors}, are all numbers in the range 0 to 15, except
for the values 5, 10, and 15 that do not occur; the number 0 denotes
stored length, while the other 12 numbers denote bonds to the twelve
nearest-neighbor sites. For every combination $(D_i, D_j)$ of those bit
patterns, we have pre-computed the lists of all bit patterns $D'_i$ and
$D'_j$ after a sideways move. Depending on the combination $(D_i, D_j)$,
at most four different sideways moves can be proposed. We therefore have
precomputed four such lists for $D'_i$ and for $D'_j$. In one step, we
first select randomly the monomer number $i$ and the list number $k$;
then we attempt a sideways move of monomer $i$ to the position
determined by the $k^{\mathrm{th}}$ list; finally, if this move does not
lead to overlapping monomers, it is accepted. This check for overlap
requires computing the position of monomers, which requires computing
the distance to the nearest tracked monomer. In our implementation, in
which this distance is at most 15 monomers, the total cpu-time required
per such step equals 90~ns.

   Not all moves are attempted with the same frequency. Since the
long-time dynamics is determined by reptation, as argued by de Gennes,
the time scale is set such that reptation moves in the interior of the
polymers are attempted with unit rate. Sideways moves which do not
increase the amount of stored length are attempted with some rate $r_s$;
the most natural choice for this rate would be unit rate once more, but
because of the much higher computational cost for these moves, we often
chose some value of $r_s<1$. As discussed above, consistency with the
density of stored length of the projected repton model demands that
moves in which the amount of stored length is increased are attempted
with a rate of $2 r_s/z$. Since the most mobile monomers hop away to
other sites with a total rate of $2 r_s$, and since 16 monomers are
potentially moved in each step, $r_s/8$ steps as described above should
be performed per monomer and per unit of time.

   Moves in which the first or last monomer joins its neighbor along the
chain can be viewed partly as reptation moves, partly as sideways
movement. On these grounds, we have decided to attempt these moves with
rate $1+r_s$. Consistency with respect to the density of stored length
requires that moves in which the first or last monomer leaves its
neighbor along the chain in a specific direction are attempted with a
rate of $2 (1+r_s)/z$.

 \section{Application: phase separation of a binary polymer mixture}

   To illustrate the efficiency of the above computational approach, we
performed a simulation of the phase separation of a binary polymer
mixture with polymer types $A$ and $B$. The $A$ and $B$ polymers
interact with a short-range repulsion, described by the Hamiltonian
 \begin{equation}
 H=J \sum_{\langle \vec{r}, \vec{r}'\rangle}
   \delta(\sigma_{\vec{r}},A)~\delta(\sigma_{\vec{r}'},B)+
   \delta(\sigma_{\vec{r}},B)~\delta(\sigma_{\vec{r}'},A)\mbox{,}
 \end{equation}
 where the summation runs over all pairs of nearest-neighbor sites, and
$\sigma(\vec{r})$ is $A$, $B$, or 0 if site $\vec{r}$ is occupied by a
polymer of type $A$, $B$ or empty, respectively. The repulsion between
$A$ and $B$ polymers provides the driving force for the phase
separation.

\begin{figure}
 \includegraphics[width=3.25in]{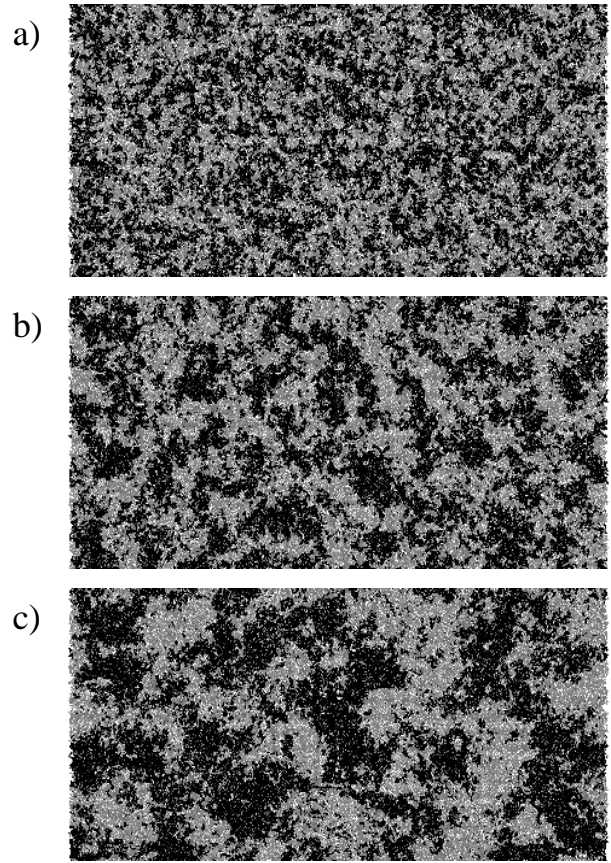}
 \caption{Two-dimensional snapshot of a three-dimensional simulation of
a binary polymer mixture with $46,080$ polymers of length $L = 100$ on a
lattice of $N = 13,824,000$ sites, at inverse temperature $\beta J =
0.1$. Snapshot a) is taken at $t=0$, i.e., the equilibrated system at
infinite temperature taken as the initial configuration. Snapshots b)
and c) are taken at $t=4.7\cdot 10^5$ and $3.7\cdot 10^6$ MC time
units.}
 \label{fig:sample}
 \end{figure}

   We simulated a system containing in total $46,080$ polymers of length
$L = 100$ on a lattice of $N = 13,824,000$ sites, at inverse temperature
$\beta J = 0.1$. The system evolves in time through reptation moves, at
unit rate, in combination with sideways moves at a rate of $r_s = 1/30$.
Figure~\ref{fig:sample} shows two-dimensional slices of the
three-dimensional system at times $t=0$, $t=4.7\cdot 10^5$, and
$t=3.7\cdot 10^6$ MC time units.

   At various times, we determine the two-point correlation function
 \begin{equation}
 g_{AB}(\vec{r})=\frac{\frac{1}{N}
      \sum_{\vec{r}'}
         \delta(\sigma_{\vec{r}'},A) ~
         \delta(\sigma_{\vec{r}'+\vec{r}},B) }
      {\left(\frac{1}{N}\sum_{\vec{r}'}
            \delta(\sigma_{\vec{r}'},A) \right) ~
         \left(\frac{1}{N}\sum_{\vec{r}'}
               \delta(\sigma_{\vec{r}'},B) \right) }\mbox{.}
 \end{equation}
 From this function we determine the spherically-averaged radial
distribution function, defined as $RDF(r) = 1-g(\vec{r})$. This function
is 1 at $r=0$, then decreases, and eventually approaches 0 for large
$r$. After some time, the conserved dynamics gives rise to damped
oscillations in the RDF. The frequency of these oscillations can be
determined from the shortest distance $r_0$ at which the RDF equals
zero. The typical domain size $d(t)$ is then obtained as twice this
distance.

\begin{figure}
 \includegraphics[width=3.25in]{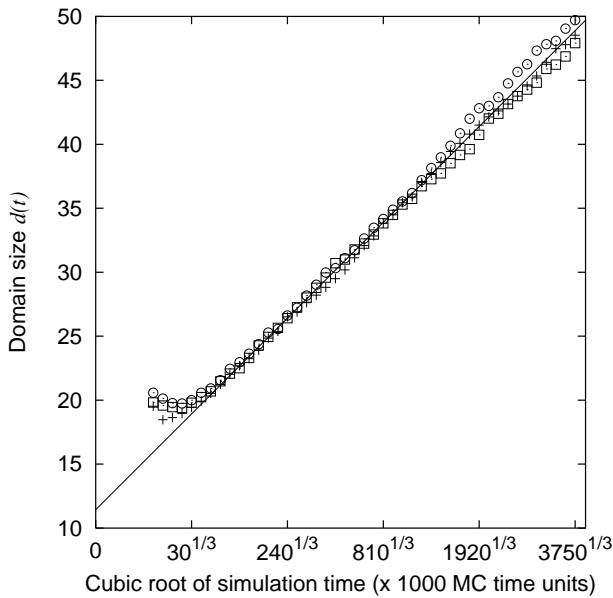}
 \caption{Domain size of a phase-separating binary polymer mixture as
function of the cubic root of the simulation time. The three symbols
indicate three independent simulations. A good agreement with domain
growth $d(t) \sim t^{1/3}$ is found.}
 \label{fig:domainsize}
 \end{figure}

   Figure~\ref{fig:domainsize}
 shows the domain size of the phase-separating mixture at different
times. The domain size $d(t)$, as defined above, is plotted as a
function of the cubic root of the simulation time in Monte Carlo sweeps.
The straight line shows that the domain size grows with $d(t) \sim
t^{1/3}$. This is to be expected in a system with conserved order
dynamics and without hydrodynamics, as in ``Model~B''~\cite{hohenberg77}.

 \section{Summary and outlook}

   In summary, we have presented a lattice model to study the dynamics
of polymeric systems. Compared to other lattice polymer models in
current use, this model lends itself for highly efficient simulation:
employing multispin coding, the computer time required per elementary
move is reduced by more than three orders of magnitude, to a few
nanoseconds.

   To demonstrate the strength of our model, we simulated the phase
separation of a binary mixture of polymers. We performed simulations of
a system with about $50,000$ polymers, each containing 100 monomers,
located on a lattice with about 14 million sites. We verified that over
the final two decades in time, the domain size $d(t)$ grows according to
$d(t) \sim t^{1/3}$, as expected for a system with overdamped dynamics
and a local conservation law. This simulation involves $3.4 \cdot
10^{13}$ elementary moves, but could nevertheless be carried out on a
single-processor workstation in about 20 days.

   Recently, we have used this lattice polymer model to study
fractionation in quasi-binary (polydisperse) polymer
mixtures~\cite{heukelum2003}. Currently, we are using the model
presented here to study the dynamics of absorption of a polydisperse
polymer mixture on a surface, as well as the sieving process of
polydisperse polymer mixtures through nanopores.

 \bibliography{polymermodel}

 \end{document}